\documentclass[prl,twocolumn,nofootinbib,superscriptaddress]{revtex4}
\usepackage{graphicx}
\usepackage{amssymb}

\usepackage{amsmath,amssymb,amsfonts,times,graphicx}
\usepackage[ps2pdf,bookmarks=true,colorlinks,linkcolor=red,urlcolor=blue,citecolor=blue]{hyperref}
\usepackage[usenames]{color}
\usepackage{dcolumn}

\begin{document}
\title{A geometric proof of the equality between entanglement and edge spectra}
\author{Brian Swingle}
\affiliation{Department of Physics, Massachusetts Institute of Technology, Cambridge, MA 02139}\affiliation{Department of Physics, Harvard University, Cambridge, MA 02138}
\author{T. Senthil}
\affiliation{Department of Physics, Massachusetts Institute of Technology, Cambridge, MA 02139}
\begin{abstract}
The bulk-edge correspondence for topological quantum liquids states that the spectrum of the reduced density matrix of a large subregion reproduces the thermal spectrum of a physical edge.  This correspondence suggests an intricate connection between ground state entanglement and physical edge dynamics.  We give a simple geometric proof of the bulk-edge correspondence for a wide variety of physical systems.  Our unified proof relies on geometric techniques available in Lorentz invariant and conformally invariant quantum field theories.  These methods were originally developed in part to understand the physics of black holes, and we now apply them to determine the local structure of entanglement in quantum many-body systems.
\end{abstract}
\maketitle

\section{Introduction}
Since the discovery of the fractional quantum Hall fluids \onlinecite{fqhe} and the subsequent elucidation of their topological structure \onlinecite{toporder1,toporder2}, it has become clear that entanglement plays a crucial role in wide variety of zero temperature quantum phases of matter.  To describe such phases of matter, it is necessary to understand their pattern of long range entanglement since they fail to be distinguished by any symmetry breaking pattern. The basic measure of entanglement in the ground state is the entanglement entropy of a subsystem defined as the von Neumann entropy of the reduced density matrix of the subsystem.  The boundary law for entanglement entropy states that the entropy of a subsystem of linear size $L$ in the ground state typically scales like $L^{d-1}$ i.e. like the boundary of the region ($d$ is the space dimension).  Ref. \onlinecite{arealaw1} provides a comprehensive review of this fundamental result.  Entanglement entropy has since been used in Refs. \onlinecite{topent1,topent2} to give meaning to the notion of long range entanglement in topological fluids.  Entanglement considerations have also led to a revolution in our understanding of 1d physics \onlinecite{dmrg_mps,vidal,time_mps_1d}, promising variational states in higher dimensions \onlinecite{peps,terg,vidal_mera}, and a classification of 1d phases \onlinecite{mps_classify1,mps_classify2}.  At present, we are still continuing to explore the role of entanglement in quantum many-body physics.

More recently, Li and Haldane in Ref. \onlinecite{ent_spec} drew attention to interesting physical information encoded in the full spectrum of the reduced density matrix of which the entanglement entropy is a single measure.  Of course, the full spectrum is complicated and for a large enough subsystem, impossible even to list effectively, but what Li and Haldane argued was that the entanglement spectrum contained a universal part, even at moderate size, that was characteristic of the phase of interest.  To understand their claim, we write the reduced density matrix of region $R$ as $\rho_R = e^{- H_R}$, where we have defined the ``entanglement hamiltonian" $H_R$.  This statement by itself contains very little information.  The boundary law implies that $H_R$ behaves, at a very crude level, like the Hamiltonian of a lower dimensional system.  What Li and Haldane argued was that for quantum hall systems the universal part of $H_R$ was actually given by the true dynamical Hamiltonian of a physical edge.  Thus the entanglement cut becomes a physical cut, and the ground state informs us about quantum dynamics at the edge.  The entanglement spectrum has since been studied intensively, see Refs. \onlinecite{espec1,espec2,espec3}. We should also mention that information equivalent to the Li-Haldane proposal can be extracted from generalizations of the von Neumann entropy called Renyi entropies which have been studied previously.  These entropies are defined as $S_{\alpha}(R) = \frac{1}{1-\alpha} \ln{\left(\mbox{tr}(\rho^{\alpha})\right)}$ and can be thought of as computing the entanglement entropy as a function of ``entanglement temperature" $1/\alpha$.

Since the Li-Haldane proposal, known as a bulk-edge correspondence, Refs. \onlinecite{edge_ent_proof1,edge_ent_proof2,edge_ent_proof3,edge_ent_proof5,edge_ent_proof4} have offered a variety of proofs of the conjecture and variations of it in various situations ranging from non-interacting topological insulators to fractional quantum Hall fluids in $2+1$ dimensions.  There were even hints of a bulk-edge correspondence in the early work of Ref. \onlinecite{topent1}.  In this paper, we present a proof of the bulk-edge correspondence for a wide variety of physical systems.  Our proof has the advantage of simplicity, physical transparency, and generality.  It is also appealingly geometrical in nature.  Our main technical tools are a set of powerful geometrical constructions in Lorentz invariant and conformally invariant field theories that relate the entanglement spectrum of special subsystems to thermal spectra in appropriate spacetimes.

We make use of existing techniques in conformal field theory described in Refs. \onlinecite{rindler_ham,sphere_ee,holoee_deriv} to rigorously establish the intuition that the boundary used to define entanglement entropy functions much like a real boundary in topological liquids.  In particular, if at a real boundary the system has gapless edge modes, then we argue that the entanglement spectrum also contains signatures of these gapless modes.  Thus we establish very generally a powerful link between entanglement and quantum dynamics, connecting ground state properties with edge dynamics.  Our results treat systems in a variety of dimensions with and without interactions in a completely unified framework.  We provide a new derivation of topological entanglement entropy in $2+1$ dimensions, a proof of the bulk-edge correspondence for entanglement spectra in fractional quantum Hall states, a proof of the bulk-edge correspondence for topological insulators in $3+1$ dimensions, and a proof of the bulk-edge correspondence for fractionalized topological insulators in $3+1$ dimensions.

This paper is organized as follows.  The first section contains an introduction and motivation.  Section two describes our geometric tools and the intuition behind them. Section three treats the original Li-Haldane situation of fractional quantum Hall fluids followed by analogous calculations for topological band insulators and a class of recently proposed fractional topological insulators.  Finally, we conclude with some comments on future directions.

\section{Entanglement Hamiltonians from geometric flows}
Our goal is to understand the spectrum of the reduced density matrix of a disk (or more generally, a ball) in various kinds of topological quantum liquids.  Our main tool is a result from relativistic conformal field theory that identifies the spectrum of the ``entanglement hamiltonian" or ``modular hamiltonian" (as it is known in the field theory literature) of a ball in $d$ spatial dimensions with the spectrum of the usual dynamical hamiltonian of the conformal field theory on a hyperbolic spacetime of the form $H^d\times R$.  As a geometric aside, for $d=2$, $H^2$ is the familiar Poincare disk which can be covered by coordinates $x,y$ with $x^2+y^2 < 1$ and has the metric
\begin{equation}
ds^2 = \frac{dx^2 + dy^2}{(1 - (x^2 + y^2))^2}.
\end{equation}
Notice that this metric is conformally equivalent to the flat metric on the unit disk i.e. $ds^2_{H^2} = \Omega^2(x,y) ds^2_{\mbox{unit disk}}$.  The boundary $x^2+y^2=1$ is a conformal boundary in that it is infinitely far away as measured in the hyperbolic metric.  We will understand what role this space plays shortly.  But before moving on, let us remember that we are interested in gapped topological liquids described by topological quantum field theories.  We are perhaps not accustomed to thinking of these objects as conformal field theories, but while they have a much larger invariance group, they certainly belong to the more general class of conformal field theories.  We will now build up to this result from a simpler result for the entanglement Hamiltonian of a half space.

It has been known for some time, motivated by studies of Unruh radiation and Hawking radiation, that the spectrum of the reduced density matrix of a half space in a relativistic quantum field theory is related to that of the Lorentz boost generator that preserves the so-called Rindler wedge (see Ref. \onlinecite{rindler_ham}). Focusing on the two dimensions $x$ and $t$ mixed by the boost, the Rindler wedge is the region of spacetime with $x\geq 0$, $x \leq t$, and $x \geq -t$.  Considering the lightcone coordinates $x^+ = t+x$ and $x^- = t- x$, boosts by velocity $v$ in the $x$ direction take $x^{\pm} \rightarrow e^{\pm \lambda} x^{\pm}$ (with $e^\lambda = \sqrt{\frac{1+v}{1-v}}$), and since the boundaries of the Rindler wedge are $x^+ = 0$ and $x^- = 0$, we see immediately the boost only moves points around within the Rindler wedge.  An alternate characterization of the Rindler wedge is simply the region of spacetime where the physics is totally controlled by the state of the half space $x > 0$ at a fixed time, it is the ``causal development" $D$ of the half space.  In fact, the reduced density matrix of the half space is nothing but a thermal state where the Hamiltonian is the generator of Rindler boosts as shown in Ref. \onlinecite{rindler_ham}. This claim can be proved using a simple path integral argument involving the Euclidean version of the Rindler boost as shown in Ref. \onlinecite{unruh_pi}, but it is already made plausible by noting that the map $\lambda \rightarrow \lambda + 2 \pi i$ leaves the coordinates unchanged.  This is consistent with a quantum state that is thermal with respect to the generator of the boost.

Returning to our original comments, this observation accounts for the radiation seen by accelerated observers, since their effective time evolution is generated by the Rindler boost.  The Minkowski vacuum looks like a thermal state for the Rindler boost generator, and accelerated observers experience such a state as an ordinary thermal bath with respect to their internal clock.  The boost generator can be written explicitly as $K \sim x \partial_t + t \partial_x$, and we easily check that $e^{\lambda K} x^{\pm} = e^{\pm \lambda} x^{\pm}$.  The operator version of this generator acting on degrees of freedom in the half space is $K = \int_{x > 0} x \mathcal{H} - t \mathcal{P}$ where $\mathcal{H}$ and $\mathcal{P}$ are the energy and momentum density respectively.  Evaluating this generator on the $t=0$ time slice provides an immediate connection between the entanglement Hamiltonian $K$ and the physical energy density $\mathcal{H}$.  The only subtlety is the question of boundary conditions addressed below.  

Remarkably, the entanglement Hamiltonian is the generator of a geometric flow in spacetime, and this flow may be interpreted as time evolution in Rindler space.  The reduced density matrix of the half space is then a simple thermal state with respect to time evolution in Rindler space.  If we change coordinates to $x = a^{-1} e^\rho \cosh{a \eta}$ and $t = a^{-1} e^\rho \sinh{a \eta}$, then the metric on the Rindler wedge takes the form $ds^2 = e^{2 \rho} (- d \eta^2 + d \rho^2)$ and the curve $\rho = 0$ has constant acceleration $a$.  The spectrum of the entanglement Hamiltonian can then be found directly in the continuum by quantizing the low energy theory in this spacetime.  A cutoff is necessary for large negative $\rho$ (otherwise the effective temperature, given by $1/\sqrt{g_{\eta \eta}}$, diverges as $\rho \rightarrow -\infty$), a fact familiar from the study of black hole thermodynamics (the Rindler spacetime spacetime approximates the near horizon limit of a black hole).

The results for the Rindler wedge are already sufficient to prove essentially everything we want, but to work with compact subsystems we must introduce a little more technology. If we further restrict ourselves to a conformal field theory, then we have an additional result about the reduced density matrix of the $d$-ball as described in Refs. \onlinecite{sphere_ee,holoee_deriv}.  Those authors showed that the spectrum of the reduced density matrix of a disk is related to the spectrum of the conventional Hamiltonian of the conformal field theory but defined on the hyperbolic space $H^d\times R$.  The proof is very similar to Rindler wedge result, and actually follows from the result for the Rindler wedge by a conformal transformation.  Ref. \onlinecite{holoee_deriv} contains details of the geometrical flow in the causal development $D$ that is mapped to ordinary time evolution in the hyperbolic space.  The precise relation is that the spectrum of the disk density matrix is the same as the spectrum of a thermal state of the conformal field theory defined on the hyperbolic space.  The radius of the original disk sets the curvature scale of the hyperbolic space and the temperature of the thermal ensemble on that space.

A very important check of this relation is that it correctly reproduces the divergences inherent in continuum entanglement entropy.  In the hyperbolic setting, infinities arise because the hyperbolic space is non-compact.  Hence the hyperbolic thermal entropy, which is the entanglement entropy, is infrared divergent due to the infinite volume.  Cutting off the hyperbolic space at large but finite size cures the divergence.  As the explicit transformation shows, this large distance cutoff of the hyperbolic space is the same as the short distance cutoff in the field theory that renders the entanglement entropy finite.  This is because the conformal boundary of the hyperbolic space is mapped to the physical boundary of the original disk whose entanglement entropy we are interested in.  The important message is that, although formally the hyperbolic space is infinite, we must always provide a cutoff of that space at large distances.  Thus we introduce a boundary into the space, and to compute the partition function properly, we must establish boundary conditions for the fields there.  For example, boundary conditions must be considered to insure that the partition function is gauge invariant.

A useful related construction allows us to obtain the entanglement entropy of the disk (or $2$-ball $B^2$) directly in terms of a partition function of the theory on a Euclidean sphere $S^3$.  This result is also proven in Ref. \onlinecite{holoee_deriv}.  The proof proceeds as in the previous case by a judicious choice of coordinate transformation and conformal transformation which maps the causal development of the disk into a spacetime whose Euclidean section is a sphere.  Now the short distance divergence of the path integral on the sphere corresponds to the short distance singularities of the entanglement entropy.  Since $\log{Z}$ gives the free energy, and since we will be interested in topological phases where the Hamiltonian is zero, the free energy directly determines the entropy (although in general one must be more careful as shown in Ref. \onlinecite{holoee_deriv}).  Before we proceed to give more intuition for these results, let us note that they have already been used to great effect in the holographic setting and in the study of subleading terms in the entanglement entropy of gapless systems.  These methods are also quite powerful in that they rigorously establish a boundary law for certain interacting gapless systems in more than one dimension.

To develop an appreciation for these tools, let us give a few more details.  We focus on the transformation to hyperbolic space since it is relevant for a finite subsystem, the disk.  Also, we will describe the method in the context of $2+1$ dimensions, but later we will use the $3+1$ dimensional version as well (the result is completely general).  Thus consider a disk, call it $A$, of radius $R$ at fixed time $t=0$ in a conformal field theory in $2+1$ dimensions.  The reduced density matrix of the disk determines the physics within a spacetime domain called the causal development $D$ of the disk.  The boundary of this domain in spacetime has the form of two cones, one opening up and one opening down, that intersect at the disk at $t=0$ as shown in Fig. 1. These cones are light sheets that delimit the regions of spacetime reachable only by signals starting in the disk at time $t=0$.  In terms of coordinates where the flat space metric is $ds^2 = -dt^2 + dr^2 + r^2 d\Omega^2_{d-1}$ and the disk is $r \leq R$, the coordinate transformation which maps the causal development of the disk to the hyperbolic space is given by
\begin{equation}
t = R \frac{\sinh{(\tau/R)}}{\cosh{u}+\cosh{(\tau/R)}} \\
r = R \frac{\sinh{u}}{\cosh{u}+\cosh{(\tau/R)}}.
\end{equation}
In terms of these coordinates, the metric reads
\begin{equation}
ds^2 = \Omega^2 \left( - d \tau^2 + R^2 (du^2 + \sinh^2{u} d\Omega^2_{d-1}) \right),
\end{equation}
where the conformal factor $\Omega^2$ is $(\cosh{u}+\cosh{(\tau/R)})^{-1}$.  Now we simply remove the conformal factor via a conformal transformation \footnote{Ref. \onlinecite{holoee_deriv} contains a minor typo: the vector $C^\mu$ defining their conformal transformation should be $(0,-1/(2R),0,0)$.}, and we immediately have the conformal field theory in hyperbolic space.  We can easily check that $r \rightarrow R$ requires $u\rightarrow \infty$.  Using the same kinds of geometric arguments that we outlined for the Rindler wedge, one can show that the state of the conformal field theory on this spacetime is thermal under the Hamiltonian $\partial_{\tau}$ with inverse temperature $2 \pi R$.

\begin{figure}
\begin{center}
\includegraphics[width=.48\textwidth]{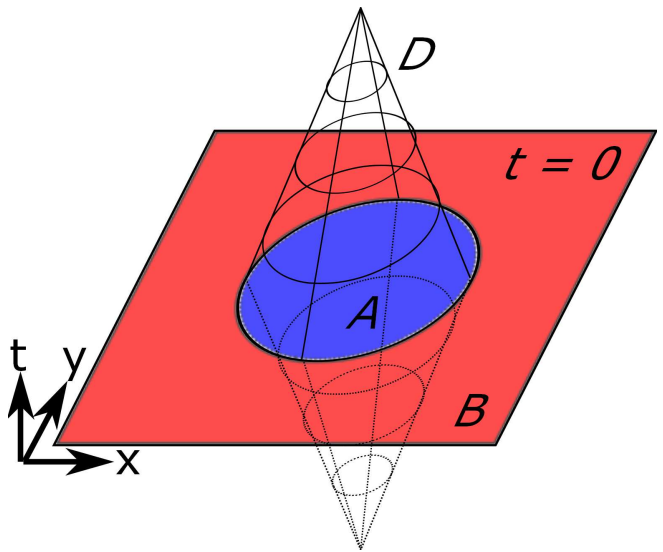}
\end{center}
\caption{The disk $A$ is colored blue.  The rest of the system at fixed time is colored red and denoted $B$.  The boundary causal development $D$ is depicted with the wire outline cones; the solid cone is the forward boundary while the dotted cone is the past boundary.  Everything inside the causal development $D$ is determined by the state on $A$ at time $t=0$ since no other region of spacetime can communicate with $D$.  Figure adapted from Ref. \onlinecite{holoee_deriv}.}
\end{figure}

To summarize, we introduced three powerful tools that provide access to the reduced density matrix of special subsystems in terms of generators of geometric flows.  First, the entanglement Hamiltonian of a half space in any relativistic field theory is given by a certain boost generator.  Second, the entanglement Hamiltonian of a ball in any conformal field theory is given by the generator of time translations in hyperbolic space.  Third, the entanglement entropy of a ball in any conformal field theory is given by the partition function of the Euclidean theory on $S^{d+1}$.  Since we consider primarily topological theories in this paper, all three results apply, but note that the half space result is especially general since it requires only Lorentz invariance.
\subsection{Warmup: topological entanglement entropy}

The simplest calculation that illustrates these tools is a computation of the entanglement entropy of a disk in topological liquids in $2+1$ dimensions.  As we already noted, topological field theories are certainly conformal field theories since they do not care about the metric of spacetime at all.  To compute $Z(S^3)$ we begin with the fact that $Z(S^2\times S^1) = 1$.  This is the statement that a topological phase has a unique ground state on the sphere.  The spacetime $S^2\times S^1$ may be cut open along the $S^2$ to yield two copies of the solid torus $B^2 \times S^1$.  Since $\partial B^2 \times S^1 = S^1 \times S^1$, $Z(B^2\times S^1) = |\Psi \rangle$ is a state in the Hilbert space of the torus generated by imaginary time evolution.  This state is normalized since $1 = Z(S^2\times S^1) = \langle \Psi | \Psi \rangle$.  Instead of gluing the tori back together directly, we first make an $S$ modular transformation of one of the boundary tori which exchanges the two non-contractible surface loops.  Now gluing the tori together yields $S^3$ as shown in Fig. 2. The modular transformation is implemented using the modular $S$-matrix $\mathcal{S}^a_b$, and a simple calculation gives $Z(S^3) = \langle \Psi | \mathcal{S} |\Psi \rangle = \mathcal{S}^0_0$.  Since $S^0_0 = 1/\mathcal{D}$ and using the fact that the ordinary Hamiltonian of a topological phase is zero, we conclude that $S(B^2) = \ln{Z(S^3)} = - \ln{\mathcal{D}}$ (the non-universal part has effectively been subtracted away) as originally shown in Refs. \onlinecite{topent1,topent2}.  In fact, using the exponentially fast factorization of the density matrix for widely separated regions and the topological invariance of $S_{\mbox{topo}}$, we may immediately establish that $S_{\mbox{topo}} = - n_{\partial} \ln{\mathcal{D}}$ where $n_{\partial}$ is the number of boundaries of the region of interest.

\begin{figure}
\begin{center}
\includegraphics[width=.48\textwidth]{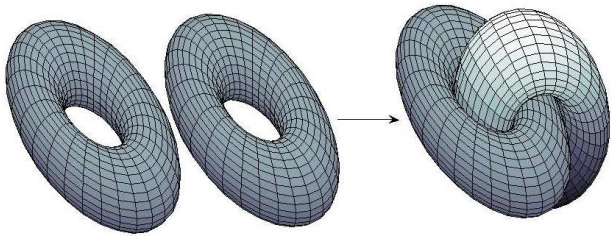}
\end{center}
\caption{Visualizing the modular transformation that maps two copies of the solid torus $B^2 \times S^1$ to the pair of interlocking solid tori on the right.  Expanding the interlocking tori to fill all of space and gluing them along their mutual boundary produces $S^3$.}
\end{figure}

\section{Bulk-edge correspondence}
\subsection{Fractional quantum hall fluids}
Now we apply the technology introduced in the previous section the problem of the entanglement spectrum in fractional quantum Hall fluids.  Consider the simplest class of topological fluids at filling fraction $\nu = 1/m$ as described by Laughlin in Ref. \onlinecite{laughlin_wf}.  These states are described at low energy by an effective Chern-Simons theory for an emergent gauge field that can be used to compute ground state degeneracy and quasiparticle statistics (see Ref. \onlinecite{Xiao-GangWenbook}.  We will now establish that the entanglement spectrum of a disk in this class of topological fluids is given by the thermal spectrum of a $1+1$ dimensional gapless edge mode.  The argument is extremely simple.

First, we map the reduced density matrix of the disk onto the thermal density matrix of the topological fluid living on hyperbolic space.  To be completely clear, we recall that only the the space is hyperbolic, time is still an extra product dimension.  Second, if the radius of curvature is small, that is, if we asked about the entanglement entropy of a large disk, then the topological fluid does not care that it lives on curved space instead of flat space.  However, there is one aspect of hyperbolic space that the fluid does care about: the existence of a (conformal) boundary.  As far as the Chern-Simons theory is concerned, the fact that this boundary is infinitely far away according to the hyperbolic metric is irrelevant, and we may as well study the Chern-Simons theory on a disk.  More practically, we cut off the hyperbolic space at some fixed size and impose boundary conditions to regularize the path integral.  As is well known, Chern-Simons theory on a manifold with boundary is not gauge invariant unless we add extra boundary degrees freedom \cite{jones_tqft}.  Because the original electron model was gauge invariant, the reduced density matrix must also be, and thus we must include these edge degrees of freedom.  But now we know the story very well, the edge degrees of freedom are those of a $c=1$ gapless boson in $1+1$ dimensions, and because the bulk is still fully gapped, these edge modes dominate the thermal physics.  Hence the ``low energy" part of the entanglement spectrum is simply given by the thermal spectrum of the corresponding $1+1$ dimensional conformal field theory.  Of course, we can similarly conclude that the entanglement spectrum of a half space in the Chern-Simons theory is given by the thermal spectrum of the same conformal field theory on an infinite line.

Note also that we are free to cut off the hyperbolic space at very large size, much larger than the curvature scale, and this means that the spectrum of the edge theory can be arbitrarily closely spaced since the spacing is set by the radius of the edge.  For a particular microscopic model, we would have to cut off the hyperbolic space at a size set (via a conformal transformation) by the physical cutoff.  Thus the details of the entanglement spectrum can depend on the microscopic cutoff and disk radius, but the existence of the gapless edge mode in the entanglement spectrum is robust.  Finally, we observe that the Renyi entropies are also easy to compute given the full spectrum.  Conformal invariance fixes the form of the leading entanglement temperature dependence, and the subleading terms reproduce the findings of Ref. \onlinecite{renyi2} that the topological part of $S_{\alpha}$ is independent of $\alpha$.

\subsection{Topological band insulators}
These techniques can also be applied to topological insulators.  Of course, for Chern insulators in $2+1$ dimensions, the result is essentially identical to the quantum Hall case.  As an illustration of the method, we directly calculate the spectrum of a lattice version of the Rindler boost Hamiltonian.  We take the lattice model in Ref. \onlinecite{2d_dirac_ham} whose low energy limit is massless Dirac fermions in 2d, and we approximate our expression for the boost Hamiltonian by multiplying the hoppings of that model with a linearly increasing function of the form $a + r_x/R$ where $a$ is analogous to the cutoff we mentioned above in Rindler space, and $r_x$ labels lattice sites in the $x$-direction.  We cutoff this model at large $r_x$ to have a finite Hamiltonian, so there is also an edge mode coming from the large $r_x$ boundary (this edge mode would be removed in a calculation of the spectra for the disk).  The model remains translation invariant in the $y$-direction.  We plot the spectra of this model in Fig. 3.

\begin{figure}
\begin{center}
\includegraphics[width=.48\textwidth]{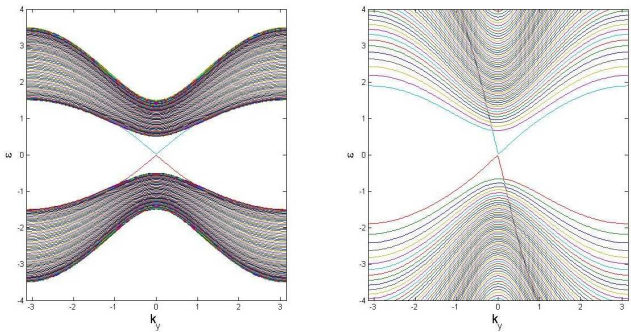}
\end{center}
\caption{Left: physical edge spectrum of the model in Ref. \onlinecite{2d_dirac_ham} with $a=1$ and $R=\infty$.  Right: spectrum of the approximate entanglement Hamiltonian with $a=1$ and $R=20$ (the spectrum is not too sensitive to the details of $a$ and $R$ and the overall scale is irrelevant).  Note that we show both spectra on the same scale, but the spectrum of the entanglement Hamiltonian is spread over a much larger range to the linearly growing couplings.  This also explains the difference in slopes for the edge modes as the edge at large $r_x$ has much larger couplings and hence a much larger speed.}
\end{figure}

Now consider the time reversal invariant $Z_2$ topological insulator in $3+1$ dimensions with effective theory given by the $\theta $ term $\theta \int F\wedge F$ with $\theta = 0,\pi$.  This is a rather simple field theory, but the $\theta$ term is topological, so we may apply our procedure.  We want to determine the entanglement spectrum of a ball in this system, and as usual, we map it to the effective thermal problem in hyperbolic space.  Cutting off the hyperbolic space at finite size once again leads to a physical boundary where we must specify boundary conditions in order to compute the partition function.  There are two classes of boundary conditions we can consider.  If we impose time reversal breaking (or particle number non-conserving) boundary conditions, then we will not in general have edge states.  However, because we are asking about the reduced density matrix deep inside the bulk of a time reversal invariant system, the only sensible boundary conditions are those that preserve time reversal.

We appeal to the robustness of the $\theta = \pi$ insulator to argue that if we do not break time reversal at the surface in our cutoff hyperbolic space, then we will have gapless edge states.  Furthermore, these edge states will not be seriously perturbed by the fact that our system lives not in flat space but in hyperbolic space as long as the curvature is smooth and the bulk gap persists.  Examining the precise form of the entanglement Hamiltonian for the half space, we see that it describes a system where the effective gap increases from zero to infinity as a function of $x$, and so it is obvious that this system will have gapless edge states of the usual form in the non-interacting model.  Microscopic interactions can be freely included so long as the system remains in the same phase and the low energy effective theory is unchanged.  Thus we have shown that the entanglement spectrum of a $3$-ball $B^3$ in a $Z_2$ non-trivial insulator is necessarily gapless just like the thermal spectrum of a physical edge.

A word of caution is appropriate here, since in the presence of interactions we cannot rule out the possibility of a surface phase transition.  Of course, whatever the surface physics, the $\theta$ terms tells us that it must be gapless and must reproduce the $1/2$ Hall response when time reversal is broken at the surface.  Lorentz invariance permits us to seriously constrain the theory, but we can only really argue for surface Dirac cones in a weakly coupled description.  If we have further information about the field theory, as in the standard non-interacting Dirac fermion model of topological insulators in $3+1$ dimensions, then we can be quite precise about the half space entanglement Hamiltonian (we don't even have to formally send the gap to infinity to obtain a topological and hence conformal theory).  On a final note, we could always consider strange boundary conditions, such as those that mimic the effects of disorder, but this would not be a good cutoff if the bulk preserved the relevant symmetries.  These considerations do suggest an interesting question: if the bulk of a topological insulator is disordered, could the entanglement spectrum track the change in the density of states at a physical surface due to this disorder?

\subsection{Fractional topological insulators in $3+1$d}
Precisely the same argument applies here for fractional topological insulators in $d=3$ spatial dimensions.  These insulators have a $\theta$ term in their low energy effective action with fractional $\theta/\pi$ as well as a topological $BF$ theory (see Ref. \onlinecite{topo_qc_review} for a brief definition) that describes ground state degeneracy.  In more familiar language, the low energy theory is deconfined $Z_n$ gauge theory in $3+1$ dimensions coupled to external electromagnetic fields via gapped fractionalized fermions that fill a topological band.  Since these phases have protected edge states so long as time reversal is preserved, we may again expect edge modes in the cutoff hyperbolic space or in Rindler space.  These edge modes will dominate the low energy thermal spectrum and hence the universal part of the entanglement spectrum.  The same caveats concerning the precise form of the edge modes applies here as well, but given a relativistic realization of the low energy effective theory, such as the one described in Refs. \onlinecite{fti1,fti2}, we can again be quite precise about the nature of the entanglement Hamiltonian.

We also note that the sphere construction can be used to compute the topological entanglement entropy of a ball inside a fractionalized topological insulator.  The topological entropy comes exclusively from the $Z_n$ gauge theory, and so we need to evaluate $Z_{Z_n}(S^4)$.  If we normalize the partition function so that $Z(S^3\times S^1) = 1$ (one state on the $3$-sphere), then we find that $Z(S^4) = 1/n$.  In fact, for any manifold $M$ we have $Z(M) = n^{b_1(M) - 1}$ where $b_1(M)$ is the first Betti number of $M$, the number of non-trivial loops in $M$.  This result follows, up to normalization, directly from the structure of the Euclidean theory as the coupling is taken to zero.  Using a lattice regulator, the flux through each plaquette is zero in the weak coupling limit, but there is a remaining freedom to specify the flux through all the non-contractible loops.  These configurations all occur with the same weight in the path integral and contribute a factor of $n$ for each loop.  Using $S_{\mbox{topo}} = \ln{Z} = - \ln{n}$ and setting $n=2$, this formula reproduces the known result for $3+1$ dimensional $Z_2$ gauge theory obtained in Ref. \onlinecite{ee_3d_z2}.

\section{Conclusion}
We have established the bulk-edge correspondence for a wide variety of topological quantum fluids.  This correspondence relates the spectrum of the reduced density matrix of a spatial subsystem in the bulk to the thermal spectrum of a physical edge.  The entanglement cut becomes a physical cut.  In addition to reproducing some old results in a unified framework, we have offered the first proof of the bulk-edge correspondence for fractional topological insulators.  We do not believe this result was seriously in doubt given the circumstantial evidence and intuition for the bulk-edge correspondence, but it is valuable to have a proof.  Although we have addressed a wide variety of systems, we believe that our technique has not been exhausted.  As an exact non-perturbative relationship between the entanglement spectrum of simple subsystems and the thermal spectrum in simple spacetimes, we believe this technique has much to offer the study of entanglement in quantum many-body systems.  As an example, it would be interesting to see if the ground state wavefunction can detect the full central charge of the edge or only the ``topological central charge" given by $c \mod 8$.  Another subtle point is the issue of inversion symmetry described in Ref. \onlinecite{edge_ent_proof5} which may require a careful study of the boundary conditions or even the other Rindler wedge associated with $x < 0$.

Although we only briefly mentioned it, our methods permit a calculation of the topological entanglement entropy of a ball in three dimensional topological phases.  We have few physical candidates for such phases, and their entanglement properties are more complicated (because we must deal with flux lines, see Ref. \onlinecite{topo_ee_3d} for a recent discussion), but we can reliably identify the subleading term in the entanglement entropy of a ball (and equivalent regions) from $Z(S^4)$.

Another interesting idea is that the mapping to hyperbolic space or Rindler space might be useful to numerically compute the entanglement spectrum of certain critical points.  We would need to study a lattice model which realizes the continuous quantum phase transition of interest on a lattice that mimics the appropriate geometry.  We may be limited by our ability to numerically simulate such a model, but it should be possible for some models and would give direct access to the entanglement spectrum.  On a related note, the mapping we have employed also suggests a way to compute entanglement entropies of critical theories very generally in terms of equivalent thermal problems.  We suspect this can be used to make statements about the computational complexity of calculating the entanglement entropy for such systems on a quantum computer.  Building on the work of Ref. \onlinecite{eevect}, we are in the process of computing the entanglement entropy of balls at the $O(n)$ Wilson-Fisher fixed point.  We end by noting that there are many other partitions of interest besides the spatial ones considered here (e.g. Refs. \onlinecite{espec_spins1,espec_spins2}), so there is still a great deal to understand about the entanglement spectrum in quantum many-body systems.

\textit{Acknowledgements}  BGS thanks Rob Myers and John McGreevy for many stimulating discussions on these topics.  BGS is supported by a Simons Fellowship through Harvard University.  TS is supported by grant number DMR-1005434.

\bibliography{ent_spec_top}

\end{document}